\documentclass[12pt,a4paper]{article}
\usepackage{cmap}
\usepackage[T2A]{fontenc}
\usepackage[cp1251]{inputenc}
\usepackage[russian,english]{babel}
\usepackage{amsmath,amsthm,amssymb}
\usepackage{amsopn}
\usepackage{bm} 
\usepackage{cite}
\usepackage[margin=25mm]{geometry} 

\newtheorem{lemma}{Лемма}
\newtheorem{theorem}{Теорема}

\usepackage{mathtools} 
\providecommand\given{} 
\newcommand\SetSymbol[1][]{%
	\nonscript\:#1\vert
	\allowbreak \nonscript\:	\mathopen{}}
\DeclarePairedDelimiterX\Set[1]\{\}{%
	\renewcommand\given{\SetSymbol[\delimsize]}	#1} 

\newcommand{\R}{\mathbb{R}}  
\newcommand{\N}{\mathbb{N}}  
\DeclareMathOperator{\conv}{conv}
\DeclareMathOperator{\ext}{ext}
\newcommand{\from}{\colon}
\renewcommand{\emptyset}{\varnothing} 
\newcommand{\BQP}{P_{\textup{BQP}}} 
\newcommand{\LOP}{P_{\textup{LO}}} 
\newcommand{\DCP}{P_{\textup{2cover}}} 
\newcommand{\Stable}{P_{\textup{stab}}} 

\usepackage[colorlinks, linkcolor=blue, pagecolor=blue, urlcolor=black]{hyperref}

\begin{document}
\title{Boolean quadric polytopes are faces\\ of linear ordering polytopes%
	\thanks{Supported by the~State Assignment for Research in P.G. Demidov Yaroslavl State University, 1.5768.2017/P220.}}
\author{A.N. Maksimenko\\[2ex]
\normalsize
P.G. Demidov Yaroslavl State University, 150000, Yaroslavl, Sovetskaya 14\\
\normalsize
\url{maximenko.a.n@gmail.com}}

\maketitle

\begin{abstract}
Let $\BQP(n)$ be a boolean quadric polytope, $n\in\N$, $\LOP(m)$~--- linear ordering polytope, $m\in\N$. 
It is shown that $\BQP(n)$ is linearly isomorphic to a face of $\LOP(2n)$.
\end{abstract}

\selectlanguage{russian}

\section{Введение}

Булевы квадратичные многогранники $\BQP(n)$, $n\in \N$, возникают, в первую очередь, в контексте поиска эффективных решений для задачи квадратичного 0/1-программирования, где $n$~--- число 0/1-переменных.
Изучению их свойств, а также свойств аффинно эквивалентных им многогранников разрезов в настоящее время уделяется значительное внимание.
Поисковая система scholar.google.ru сообщает о 889 цитированиях\footnote{дата обращения: 20.04.2017} монографии Деза и Лоран~\cite{Deza:1997}, целиком посвященной этой теме. 

С оптимизационной задачей о линейном упорядочивании связаны многогранники линейных порядков. Они не столь популярны, как булевы квадратичные. Тем не менее, исследованию их свойств уделяется внимание в нескольких десятках публикаций (см., например, \cite{Fiorini:2006,Kovalev:2012}, а также ссылки в них).

В настоящее время известно~\cite{Maksimenko:2016}, что булевы квадратичные многогранники аффинно эквивалентны граням многогранников, ассоциированных со следующими NP-трудными задачами комбинаторной оптимизации: задача коммивояжера, задача о рюкзаке, задачи о покрытиях и упаковках множества, задача о максимальной 3-выполнимости, задача о 3-назначениях и многие другие.
Известно также, многогранники любой линейной задачи комбинаторной оптимизации являются аффинными проекциями граней булевых квадратичных многогранников~\cite{Maksimenko:2012}.
В частности, задача о линейном упорядочивании является линейной задачей комбинаторной оптимизации (но не благодаря <<линейности>> упорядочивания).

Ниже будет показано, что между $\BQP(n)$ и многогранниками линейных порядков $\LOP(m)$, где $m$~--- число упорядочиваемых элементов, имеется более тесная связь. А именно, $\BQP(n)$ аффинно эквивалентен некоторой грани $\LOP(2n)$ для любого $n\in\N$.

\section{Основная часть}

Для множества $\{1,2,\dots,n\}$ будем пользоваться обозначением $[n]$.
Вектор"=столбцы (в частности, вершины многогранников) выделены полужирным шрифтом.
Гиперплоскость $H$ называется \emph{опорной} к многограннику $P$, если $P \cap H \neq \emptyset$ и многогранник $P$ целиком лежит с одной стороны от этой гиперплоскости.
Пересечение многогранника с одной или несколькими опорными гиперплоскостями называется его \emph{гранью}.

В настоящей работе рассматриваются только выпуклые многогранники, определяемые посредством описания множества их вершин.
Поэтому далее, с целью упрощения рассуждений, выпуклый многогранник будет отождествляться с множеством его вершин. 
То же самое верно и в отношении граней многогранников.
Выгоду от такого упрощения проиллюстрируем следующим примером.

Пусть $X = \ext(P)$~--- множество вершин многогранника $P$. Соответственно, $P = \conv(X)$. Пусть $H$~--- гиперплоскость, опорная к многограннику $P$.
Тогда множество вершин грани $P\cap H$ можно выразить как сложной формулой $\ext(\conv(X)\cap H)$, так и простой $X\cap H$. 
Таким образом, данное выше определение грани не искажается при замене многогранника множеством его вершин.

Пусть $P$ и $Q$~--- выпуклые многогранники, связанные аффинным отображением $\alpha\from P \to Q$. В этом случае будем говорить, что $Q$ является \emph{(аффинной) проекцией} $P$. Если же отображение $\alpha$ обратимо, то $P$ и $Q$ называются \emph{аффинно эквивалентными}. 

Множество вершин $\BQP(n)$ \emph{булева квадратичного многогранника} состоит из 0/1-векторов $\bm{x}=(x_{ij}) \in \{0, 1\}^{n(n+1)/2}$, координаты которых удовлетворяют условию
\begin{equation}
\label{eq:BQP}
x_{ij} = x_{ii} x_{jj}, \quad 1 \le i < j \le n.
\end{equation}
Этот многогранник обладает многими интересными свойствами.
Наиболее впечатляющим свойством является то, что размер его линейного (внешнего) описания остается экспоненциальным, даже если разрешено использование вспомогательных переменных~\cite{Fiorini:2015,Kaibel:2015}.
Его граф полон; более того, любые три вершины образуют грань этого многогранника~\cite{Deza:1997} (многогранники с таким свойством называются 3-смежностными). 
Известно также, что для любого фиксированного 
$s \le 3 \left\lfloor \frac{\log_2 n}{2}\right\rfloor$, 
$\BQP(n)$ имеет $s$-смежностную грань со сверхполиномиальным (относительно $n$) числом вершин~\cite{Maksimenko:2014}.

Перейдем теперь к описанию многогранника линейных порядков.

Пусть $D = (V, A)$ "--- орграф без петель и параллельных дуг, с множеством вершин $V = [m]$.
Далее предполагается, что орграф $D$ полный, то есть $(i,j) \in A$ и $(j,i) \in A$ для всех $i,j \in V$, $i \neq j$.
Подграф $T = (V, A')$ орграфа $D$ называется \emph{транзитивным}, если из условий $(i, j) \in A'$ и $(j, k) \in A'$ следует $(i, k) \in A'$.
Если для каждой пары вершин $i,j \in V$ в $A'$ входит ровно одна из двух дуг $(i, j)$ и $(j, i)$, то соответствующий орграф называется \emph{турниром}.
Транзитивный турнир (точнее, множество его дуг $A'$) в орграфе $D$ будем называть \emph{линейным порядком}.
Каждый линейный порядок $L$ в $D$ соответствует некоторой перестановке $\pi \from [n] \to [n]$, удовлетворяющей условию 
\begin{equation}
\label{eq:piLinear}
\pi(i) < \pi(j) \iff (i,j) \in L.
\end{equation}

Координаты $y_{ij}$, $1 \le i < j \le m$ характеристического вектора $\bm{y} \in \R^{m(m-1)/2}$ для линейного порядка $L$ в $D$ определим следующим образом:
\[
y_{ij} = \begin{cases}
1, &\text{если $(i,j)\in L$,}\\
0, &\text{если $(j,i)\in L$.}
\end{cases}
\]
Обозначим через $\LOP(m)$ множество всех характеристических векторов линейных порядков в~$D$.
Выпуклая оболочка $\LOP(m)$ называется \emph{многогранником линейных порядков}. 
$\LOP(m)$ может быть также определен как множество 0/1"~векторов $\bm{y}\in\{0,1\}^{m(m-1)/2}$, удовлетворяющих 3-контурым неравенствам (см., например, \cite{Kovalev:2012}):
\begin{equation}
\label{3cycle}
0 \le y_{ij} + y_{jk} - y_{ik} \le 1, \quad i < j < k.
\end{equation}

Рассмотрим еще два семейства многогранников.

\emph{Многогранником двойных покрытий} будем называть выпуклую оболочку множества
\[
\DCP(B) =  \Set*{\bm{x}\in\{0,1\}^n \given B \bm{x} = \bm{2}},
\]
где $B \in \{0,1\}^{k\times n}$~--- ($k\times n$)-матрица, каждая строка которой содержит ровно четыре единицы, $\bm{2}$~--- вектор"=столбец, все координаты которого равны 2.
Известно, что это семейство является в некотором смысле минимальным среди семейств многогранников с NP-полным критерием несмежности вершин~\cite{Maksimenko:2014b}.

Легко заметить, что $\LOP(m)$ аффинно эквивалентен грани $\DCP(B)$, где система $B \bm{x} = \bm{2}$ определяется следующим образом.
В дополнение к переменным $y_{ij}$, $1 \le i < j \le m$, вводятся две переменные $z,h \in \{0,1\}$.
Для каждой переменной $y_{ij}$ вводится дополнительная переменная $\bar{y}_{ij} \in \{0,1\}$ и уравнение
\begin{equation}\label{eq:DCP1}
y_{ij} + \bar{y}_{ij} + z + h = 2.
\end{equation}
А каждое 3-контурное неравенство \eqref{3cycle} заменяется уравнением
\begin{equation}\label{eq:DCP2}
y_{ij} + y_{jk} + \bar{y}_{ik} + t_{ijk} = 2,
\end{equation}
где $t_{ijk}$~--- еще одна дополнительная 0/1-переменная.
С одной стороны, система из $m(m-1)(m+1)/6$ уравнений \eqref{eq:DCP1} и \eqref{eq:DCP2} определяет некоторый многогранник двойных покрытий.
С другой стороны, в пересечении опорных гиперплоскостей $z=0$ и $h=1$
находится грань этого многогранника, аффинно эквивалентная $\LOP(m)$.
В этой связи заметим, что многогранники двойных покрытий (во всяком случае, некоторые из них) едва ли могут быть гранями $\LOP(m)$, так как критерий смежности вершин последнего полиномиален~\cite{Young:1978}.

Пусть $G=(V,E)$~--- (неориентированный) граф и $V = [n]$.
\emph{Многогранником независимых множеств} в графе $G$ называется выпуклая оболочка множества
\[
\Stable(G) = \Set*{\bm{x}\in\{0,1\}^n \given x_i + x_j \le 1 \text{ для каждого ребра } \{i,j\} \in E}.
\]
Сводимость задачи о независимом множестве к задаче о разрезающем циклы наборе дуг, описанная в классической работе Карпа~\cite{Karp:1972}, определяет следующую взаимосвязь между $\Stable(G)$ и $\LOP(m)$.

\begin{lemma}\label{lem:1}
Пусть $G = (V,E)$~--- граф, где $V = [n]$. 
Тогда многогранник $\Stable(G)$ является проекцией одной из граней многогранника $\LOP(2n)$.
\end{lemma}
\begin{proof}
Пусть $\bm{y} = (y_{ij}) \in \LOP(2n)$.
Воспользуемся тем, что уравнения \(y_{ij} = 0\), при $1 \le i < j \le 2n$ определяют (некоторые) опорные гиперплоскости многогранника $\LOP(2n)$.
Для каждого ребра $\{i, j\}\in E$ графа $G$ положим 
\begin{equation}
\label{eq:StableLOP}
y_{i, n+j} = y_{j, n+i} = 0
\end{equation}
и перейдем к рассмотрению соответствующей грани $F(G)$ многогранника $\LOP(2n)$.
Ниже будет показано, что эта грань связана с многогранником $\Stable(G)$ аффинным отображением $\alpha$:
\[
y_{i, n+i} \mapsto x_i, \quad \text{где } \bm{x} = (x_i) \in \Stable(G).
\]

Из 3-контурных неравенств 
\begin{align}
y_{i, j} + y_{j, n+j} - y_{i, n+j} &\le 1, \\
y_{i, j} + y_{j, n+i} - y_{i, n+i} &\ge 0,
\end{align}
и ограничений \eqref{eq:StableLOP}
следует $y_{i, n+i} \le y_{i, j} \le 1 - y_{j, n+j}$ и, в частности,
\[
y_{i, n+i} + y_{j, n+j} \le 1, \quad \text{при } \{i, j\}\in E.
\]
Таким образом, $\alpha(F(G)) \subseteq \Stable(G)$.

Покажем теперь, что для каждой вершины $\bm{x} \in \Stable(G)$ найдется вершина $\bm{y} \in F(G)$ такая, что $y_{i, n+i} = x_i$, $i\in[n]$.

Для произвольно выбранной вершины $\bm{x} \in \Stable(G)$ рассмотрим
множества
\[
I_0 = \Set*{i \in [n] \given x_i = 0} \quad \text{и} \quad I_1 = \Set*{i \in [n] \given x_i = 1}.
\]
Далее предполагаем, что элементы множеств $I_0 = \{i_1, \dots, i_k\}$ и $I_1 = \{i'_1, \dots, i'_{n-k}\}$ отсортированы (любым способом).
Соответствующий линейный порядок для $\bm{y} \in F(G)$ представим перестановкой $\pi \from [n] \to [n]$ (см. условие \eqref{eq:piLinear}).
Положим
\begin{align*}
\pi(i_s) &= 2n-k + s, &
\pi(n+i_s) &= s, & s &\in [k],\\
\pi(i'_t) &= k + t, &
\pi(n+i'_t) &= n + t, & t &\in [n-k].
\end{align*}
Из описания перестановки $\pi$ следует, что 
$y_{i_s, n+i_s} = 0$,  при $s \in [k]$, а
$y_{i'_t, n+i'_t} = 1$, при $t \in [n-k]$.
То есть $\alpha(\bm{y}) = \bm{x}$.
Кроме того, если $y_{i, n+i} + y_{j, n+j} \le 1$, то $y_{i, n+j} = y_{j, n+i} = 0$. Значит, $\bm{y} \in F(G)$.
\end{proof}

Заметим, что в только что доказанной лемме речь идет именно о проекции,
так как каждая вершина $\bm{x} \in \Stable(G)$, вообще говоря, является образом нескольких вершин соответствующей грани многогранника $\LOP(2n)$. 

Известно~\cite{Maksimenko:2016}, что для каждого $n\in\N$ существует граф $G$ на $n(n+1)$ вершинах такой, что булев квадратичный многогранник $\BQP(n)$ аффинно эквивалентен грани многогранника $\Stable(G)$.
С учетом леммы \ref{lem:1} получаем, что $\BQP(n)$ является проекцией одной из граней многогранника $\LOP(2n(n+1))$.
Оказывается, между этими двумя многогранниками имеется более тесная связь.

\begin{theorem}\label{thm:1}
	$\BQP(n)$ аффинно эквивалентен одной из граней многогранника $\LOP(2n)$, $n\in\N$.
\end{theorem}

Таким образом, многие свойства булева квадратичного многогранника наследуются многогранником линейных порядков.
Например, так как граф многогранника $\BQP(n)$ полон~\cite{Deza:1997} и имеет $2^n$ вершин, то кликовое число графа многогранника $\LOP(m)$ ограничено снизу величиной $2^{\lfloor m/2\rfloor}$.
Аналогично, так как сложность расширенной формулировки для $\BQP(n)$ оценивается снизу величиной $1.5^n$~\cite{Kaibel:2015}, то сложность расширенной формулировки для $\LOP(m)$ также экспоненциальна относительно $m$.

\section{Доказательство теоремы \ref{thm:1}}
Пусть $\bm{y} = (y_{ij}) \in \LOP(2n)$.
Воспользуемся тем, что неравенства $y_{ij} \ge 0$, при $1 \le i < j \le 2n$,
и 3-контурные неравенства \eqref{3cycle} выполнены для всех $\bm{y} \in \LOP(2n)$,
а соответствующие равенства определяют (некоторые) опорные гиперплоскости для многогранника $\LOP(2n)$.

Покажем, что многогранник $\BQP(n)$ аффинно эквивалентен грани $F \subset \LOP(2n)$, лежащей в пересечении опорных гиперплоскостей
\begin{align}
y_{2i, 2j-1} &= 0, \label{eq:LOP1}\\
y_{2i-1, 2i}   + y_{2i, 2j}   - y_{2i-1, 2j} &= 0, \label{eq:LOP2}\\
y_{2i-1, 2j-1} + y_{2j-1, 2j} - y_{2i-1, 2j} &= 0, \label{eq:LOP3}
\end{align}
для всех $1 \le i < j \le n$.

Из \eqref{eq:LOP2} и \eqref{eq:LOP3} выводим
\begin{align}
y_{2i-1, 2j}   &= y_{2i-1, 2i} + y_{2i, 2j}, \label{eq:LOP4}\\
y_{2i-1, 2j-1} &= y_{2i-1, 2i} + y_{2i, 2j} - y_{2j-1, 2j}. \label{eq:LOP5}
\end{align}
Таким образом, все координаты вектора $\bm{y} \in F$ линейно зависят от координат $y_{2i-1, 2i}$, $i\in[n]$, и $y_{2i, 2j}$, $1 \le i < j \le n$.

Покажем, что значения координаты $y_{2i, 2j}$ однозначно определяются значениями координат $y_{2i-1, 2i}$ и $y_{2j-1, 2j}$.
Из \eqref{eq:LOP4} и $y_{2i-1, 2j} \le 1$ следует \(y_{2i, 2j} \le 1 - y_{2i-1, 2i}\), иными словами,
\[
y_{2i-1, 2i} = 1 \Rightarrow y_{2i, 2j} = 0.
\]
Из 3-контурного неравенства $0 \le y_{2i, 2j-1} + y_{2j-1, 2j} - y_{2i, 2j}$ и уравнения \eqref{eq:LOP1} следует \(y_{2i, 2j} \le y_{2j-1, 2j}\),
то есть
\[
y_{2j-1, 2j} = 0 \Rightarrow y_{2i, 2j} = 0.
\]
А из \eqref{eq:LOP5} и $y_{2i-1, 2j-1} \ge 0$ следует \(y_{2i, 2j} \ge y_{2j-1, 2j} - y_{2i-1, 2i}\), то есть 
\[
y_{2i, 2j} = 1, \quad \text{если $y_{2i-1, 2i} = 0$  и $y_{2j-1, 2j} = 1$}.
\]
Таким образом, учитывая, что вектор $\bm{y}\in F$ является 0/1-вектором,
\begin{equation}
\label{eq:LOPbqp}
y_{2i, 2j} = y_{2j-1, 2j} (1 - y_{2i-1, 2i}).
\end{equation}

Итак, все вершины грани $F$ должны быть 0/1-векторами, удовлетворяющими соотношению \eqref{eq:LOPbqp}, и все координаты этих векторов линейно зависят от $y_{2i-1, 2i}$, $i\in[n]$, и $y_{2i, 2j}$, $1 \le i < j \le n$ (см. уравнения \eqref{eq:LOP1}, \eqref{eq:LOP4} и \eqref{eq:LOP5}).
Покажем теперь, что каждому набору значений переменных $y_{2i-1, 2i}$, $i\in[n]$,
на самом деле соответствует некоторая вершина грани $F$.

Пусть 
\[
I_0 = \Set*{i \in [n] \given y_{2i-1, 2i} = 0}, \quad I_1 = \Set*{i \in [n] \given y_{2i-1, 2i} = 1}.
\]
Далее предполагаем, что элементы множеств $I_0 = \{i_1, \dots, i_k\}$ и $I_1 = \{i'_1, \dots, i'_{n-k}\}$ отсортированы \emph{по убыванию}.
Линейный порядок для соответствующей вершины $\bm{y} \in F$ представим перестановкой $\pi \from [n] \to [n]$ (см. условие \eqref{eq:piLinear}).
Положим
\begin{align*}
\pi(2i_s-1) &= n-k + 2 s, &
\pi(2i_s) &= \pi(2i_s-1)-1, & s &\in [k],\\
\pi(2i'_t-1) &= t, &
\pi(2i'_t) &= n + k + t, & t &\in [n-k].
\end{align*}
Так, например, в случае $n=3$ вершины грани $F \subset \LOP(6)$ соответствуют восьми перестановкам (записанным в виде $\pi^{-1}(1)\ldots\pi^{-1}(6)$, то есть если цифра $i$ располагается в этой последовательности левее $j$, то $y_{ij} = 1$)
\begin{alignat*}{4}
&654321, \qquad & k &=3 , \quad & I_0 &= \{3,2,1\}, \quad& I_1 &= \emptyset,\\
&165432, & k &=2 , \quad & I_0 &= \{3,2\},   & I_1 &= \{1\},\\
&365214, & k &=2 , \quad & I_0 &= \{3,1\},   & I_1 &= \{2\},\\
&543216, & k &=2 , \quad & I_0 &= \{2,1\},   & I_1 &= \{3\},\\
&316542, & k &=1 , \quad & I_0 &= \{3\},     & I_1 &= \{2,1\},\\
&514362, & k &=1 , \quad & I_0 &= \{2\},     & I_1 &= \{3,1\},\\
&532164, & k &=1 , \quad & I_0 &= \{1\},     & I_1 &= \{3,2\},\\
&531642, & k &=0 , \quad & I_0 &= \emptyset, & I_1 &= \{3,2,1\}.
\end{alignat*}

Из описания перестановки $\pi$ следует, что 
$y_{2i_s - 1, 2i_s} = 0$,  при $s \in [k]$, а
$y_{2i'_t - 1, 2i'_t} = 1$, при $t \in [n-k]$.
Справедливость соотношений \eqref{eq:LOP1}--\eqref{eq:LOP3} проверяется перебором четырех случаев, в зависимости от принадлежности индексов $i,j$ множествам $I_0$, $I_1$.

Завершая доказательство, установим между $\bm{x} = (x_{ij}) \in \BQP(n)$ и $\bm{y} \in F \subset \LOP(2n)$ взаимно"=однозначное соответствие:
\begin{align*}
x_{ii} &= y_{2i-1, 2i}, & i &\in [n],\\
x_{ij} &= y_{2j-1, 2j} - y_{2i, 2j}, & 1 &\le i < j \le [n].
\end{align*}

\section{Благодарности}

Автор благодарит Самюэля Фиорини за то, что он указал на взаимосвязь между многогранниками независимых множеств и многогранниками линейных порядков, описанную в лемме~\ref{lem:1}.

\end{document}